# On the Bandwidth of High-Impedance Frequency Selective Surfaces

Filippo Costa, *Student Member, IEEE*, Simone Genovesi, *Member, IEEE*, and Agostino Monorchio, *Senior Member, IEEE*

*Abstract*—In this letter, the bandwidth of high-impedance surfaces (HISs) is discussed by an equivalent circuit approach. Even if these surfaces have been employed for almost 10 years, it is sometimes unclear how to choose the shape of the frequency selective surface (FSS) on the top of the grounded slab in order to achieve the largest possible bandwidth. Here, we will show that the conventional approach describing the HIS as a parallel connection between the inductance given by the grounded dielectric substrate and the capacitance of the FSS may induce inaccurate results in the determination of the operating bandwidth of the structure. Indeed, in order to derive a more complete model and to provide a more accurate estimate of the operating bandwidth, it is also necessary to introduce the series inductance of the FSS. We will present the explicit expression for defining the bandwidth of a HIS, and we will show that the reduction of the FSS inductance results in the best choice for achieving wide operating bandwidth in correspondence with a given frequency.

## I. INTRODUCTION

HIGH-IMPEDANCE surfaces (HISs) [1] have been frequently employed in electromagnetic devices, as for instance in the design of low-profile antennas [2], [3] or improved electromagnetic absorbers [4], [5]. As is well known, these structures mimic the perfect magnetic conductor (PMC) condition within a small frequency range, and for this reason, they are often referred to as artificial magnetic conductors (AMCs). It is therefore evident that the bandwidth of this metasurface is one of the key features of the structure. Some papers available in literature often present novel artificial surfaces claiming a large operational bandwidth. However, in order to make a fair comparison, structures with equal substrate should be considered since the substrate parameters are essential in defining the HIS bandwidth. In this work, we show that the conventional representation of the HIS as a parallel circuit between the inductance of the grounded dielectric and the capacitance given by the capacitive frequency selective surface (FSS) is not fully accurate for the determination of the device bandwidth. In particular, we demonstrate that it is necessary to include the series inductance of the FSS in the equivalent circuit for obtaining the correct expression of the bandwidth.

Afterward, the influence of the cell shape on the HIS bandwidth is analyzed. Simple considerations on the equivalent circuit allow us to define the shape of the periodic cell ensuring the widest frequency band. In particular, we are interested in deriving the most wideband FSS shape in correspondence with a fixed frequency since this issue has many interesting practical implications. Some examples of AMC structures, whose substrate thickness and permittivity are kept constant, are presented and discussed.

## II. FORMULATION

The equivalent input impedance of the reactive surface $Z_R$ can be interpreted as the parallel connection of the FSS impedance $Z_{FSS}$ and the equivalent input impedance of the grounded dielectric slab, $Z_d$. For normal incidence, the impedance of the (thin) grounded dielectric slab behaves as an inductor ($Z_s = j\mu d$, $\mu$ and $d$ being the permeability and thickness, respectively, of the dielectric slab). The FSS element can be modeled as a single capacitance in the quasi-static region or by a *RLC* series circuit for a larger frequency range. Considering the FSS element as a pure capacitive element $C_{FSS}$, the bandwidth $BW$ of the HIS structure, defined as the range where the phase of reflection coefficient is comprised between $+90°$ and $-90°$, can be written as the bandwidth of a parallel *LC* circuit

$$\frac{BW}{\omega_0} = \frac{1}{\zeta_0}\sqrt{\frac{L_s}{C_{FSS}}} \qquad (1)$$

where $L_s$ represents the inductance of the substrate, $\omega_0$ is the angular resonance frequency, and $\zeta_0$ is the free-space impedance. From the relation (1), it results that an increase of the substrate thickness enlarges the bandwidth of the structure due to the growth of the inductance. We also mention that the use of a magnetic substrate would be a valuable strategy for obtaining a higher inductance value and, as a consequence, a bandwidth increment [6].

However, from (1), one might conclude that, while keeping the substrate thickness and the resonance frequency constant, a highly capacitive element such as a patch with a very small gap would be characterized by a smaller bandwidth than a less coupled element as a cross. Moreover, if we model the FSS by using a capacitor only, we could speculate that different shapes of the patch, with different gaps and different periodicities but with the same value of averaged capacitance [7], would have the same bandwidth. Contrarily, these considerations are incorrect because the simple *LC* parallel circuit, even if it is an excellent tool for studying these structures [5], [7], does not allow to correctly predict the operating bandwidth.

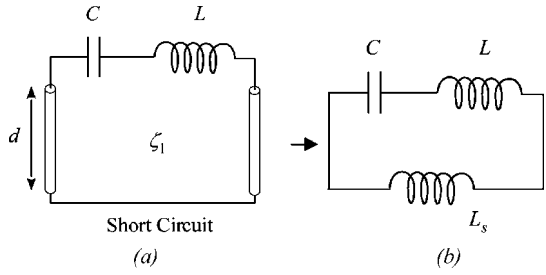

Fig. 1. Equivalent transmission line of a (a) high-impedance surface and (b) its corresponding lumped circuit model allowing to correctly determine the bandwidth of the structure.

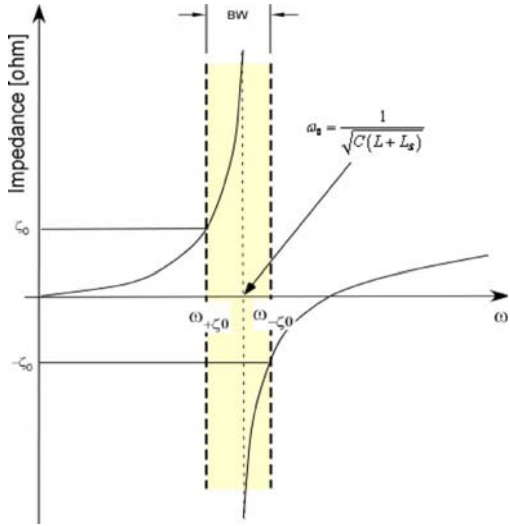

Fig. 2. Qualitative behavior of the reactive impedance in (2) and graphical representation of the HIS bandwidth.

In order to properly evaluate the bandwidth of the HIS structures, it is necessary to consider a more complete circuit where the series inductance $L$ of the FSS has been introduced (as shown in Fig. 1).

The impedance of the circuit reported in Fig. 1(b) reads

$$Z = j\frac{\omega L_s \left(1 - \omega^2 LC\right)}{1 - \omega^2 C \left(L + L_s\right)}. \quad (2)$$

As is well known, the high-impedance condition is verified when the denominator of (2) is null, namely at the angular resonance frequency $\omega_0 = 1/\sqrt{C(L + L_S)}$. The bandwidth of a high-impedance surface is defined as the frequency range where the absolute value of the impedance is larger than the free-space impedance (see Fig. 2).

This condition corresponds to a phase of reflection coefficient comprised within the range $[+90, -90°]$ [1]. In order to define the bandwidth of the new equivalent circuit, it is necessary to find the two positive frequencies for which the impedance in (2) is equal to $+\zeta_0$ and $-\zeta_0$ (see Fig. 2). In the former case, the equation has two positive solutions for $\omega$, and we must select the first one; indeed, the second one would be obtained when the HIS impedance is equal to $\zeta_0$ at higher frequency, well after the AMC band (see Fig. 2). In the latter case, the equation has only a positive solution for $\omega$. The two equations can be solved by employing the theory of the cubic equations [8]. By dividing the equations by the coefficient of $\omega^3$, we can write the coefficients in the other three terms as follows:

$$A_2 = \mp\frac{\zeta_0(L_s + L)}{L_s L}, \quad A_1 = -\frac{1}{LC}, \quad A_0 = \frac{\pm\zeta_0}{L_s LC}. \quad (3)$$

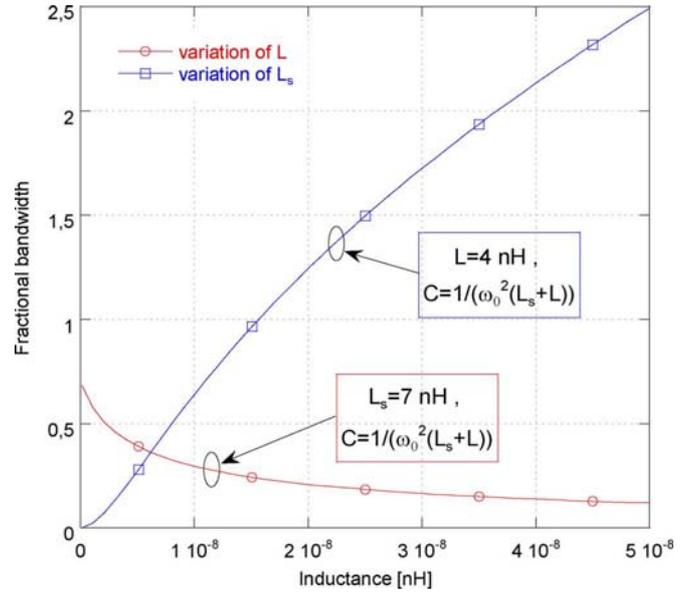

Fig. 3. Fractional bandwidth of a high-impedance surface as a function of the FSS series inductance $L$ (by fixing the grounded substrate inductance $L_s$ to 7 nH) and as a function of $L_s$ (by fixing $L$ to 4 nH). The capacitance is varied in order to preserve the same resonance frequency that is fixed to 6 GHz in all cases.

By defining the three following intermediate coefficients:

$$Q = \frac{3A_1 - A_2^2}{9}$$
$$R = \frac{9A_1 A_2 - 27A_0 - 2A_2^3}{54}$$
$$\vartheta = \cos^{-1}\left(\frac{R}{\sqrt{-Q^3}}\right) \quad (4)$$

we can get the two right solutions and, after some algebra, we can also derive the final expression of the fractional bandwidth of the new circuit ($FBW = (\omega_{-\zeta_0} - \omega_{+\zeta_0})/\omega_0$) as

$$FBW = \sqrt{C(L+L_S)}\left[\sqrt{-Q}\left(\cos(\vartheta) + \sqrt{3}\sin(\vartheta)\right) - \frac{2}{3}A_2\right]. \quad (5)$$

In order to highlight how the FSS series inductance influences the fractional bandwidth of the high-impedance surface, the relation (5) is plotted in Fig. 3 as a function of $L_S$ and $L$. When the values of the inductances are varied, the capacitance value is changed as well in order to maintain the resonance frequency fixed. In this case, the resonance frequency of the resonator is chosen to be equal to 6 GHz. The values for the inductances and the capacitance are realistic values in a practical design. As expected, increasing the grounded substrate inductance leads to an enlargement of the fractional bandwidth. Conversely, it is worth underlining that an increase of the series inductance implies a reduction of the fractional bandwidth.

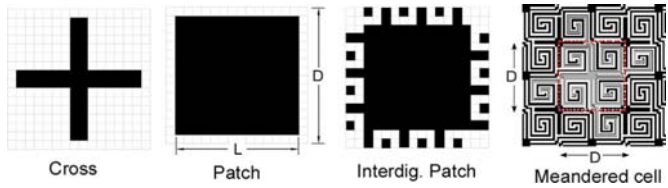

Fig. 4. FSS element shapes under investigation.

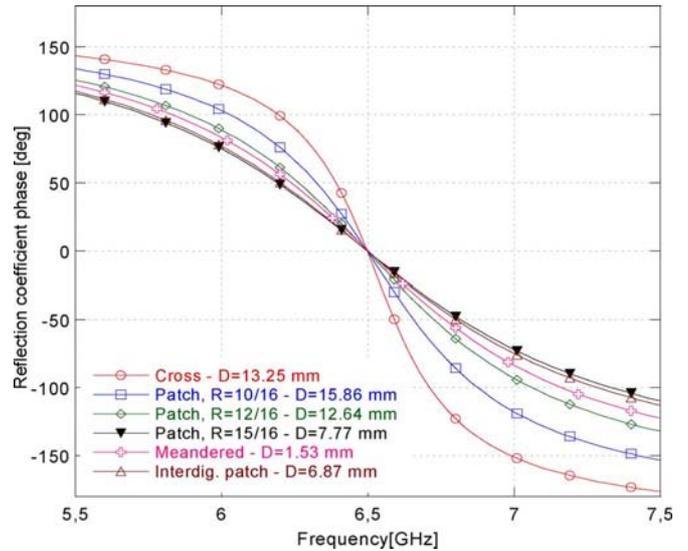

Fig. 5. Phase of reflection coefficient of different HIS on the same substrate and a periodicity ($D$) designed so that all the analyzed structures resonate at the same frequency.

The fractional bandwidth of this resonant circuit can be maximized both by decreasing the inductance and the capacitance of the FSS (considering that the inductance of the substrate is fixed). Anyway, in order to preserve the same resonance frequency, an increase of the capacitance value must correspond to a decrease of the series inductance and vice versa. As it is apparent from Fig. 3, the minimization of the inductance of the element and the consequent increase of the capacitance results in the best choice for maximizing the fractional bandwidth. The inductance of an FSS can be lowered by using large elements with respect to the unit cell (as, for instance, the solid interior elements defined in [4]). This argument can be intuitively explained by considering the inductance of a straight wire above the ground plane. According to the expression reported in [9], it follows that the inductance of a printed structure can be lowered by using large element for a constant value of substrate thickness. For these reasons, we expect that a cross element will be the most narrowband, whereas the patch-type element will be the most wideband even if its capacitance is higher.

## III. NUMERICAL RESULTS

To demonstrate these intuitive considerations, we analyze six different FSS shapes on the same grounded dielectric substrate (Fig. 4). The substrate parameters are $\varepsilon_r = 4.5 - j0.088$ and thickness equal to 1.6 mm (typical of the commercial FR4). Patch elements characterized by a different ratio $R$ between the patch size $L$ and the periodicity $D$ ($R = L/D$) are analyzed together with a cross-shaped FSS, an interdigitated patch, and a meandered shape [10], [11]. Another candidate offering large operational bandwidths is the hexagon element, but it does not guarantee better performance in terms of bandwidth with respect to the square patch.

To obtain a resonance in correspondence with the same frequency, the different elements must have different periodicities. Fig. 5 shows the phase of reflection coefficient of different HIS configurations with the same substrate and a periodicity designed so that all the analyzed structures resonate at the same frequency. To confirm the validity of our model [and therefore the bandwidth relation presented in (5)], the impedances of the FSSs under analysis were determined by the inversion method proposed in [12]. Once computed, each FSS impedance, the values of the capacitance, and the inductance can be obtained by an iterative procedure. The impedances for each FSS element employed in the HIS configurations are reported in Fig. 6 together with its approximation by the series $LC$ circuit. In Table I, a summary of the capacitances and inductances for each analyzed shape is reported together with the fractional bandwidth computed in accordance to the relation (5). From these values, it is apparent that an increase of the FSS capacitance does not correspond to a decrease of the bandwidth as one could infer by observing the relation (1). This happens because, as for instance in the patch shape, the reduction of the distance between adjacent patches implies an increase of the capacitance but, at the same time, a strong reduction of the inductance.

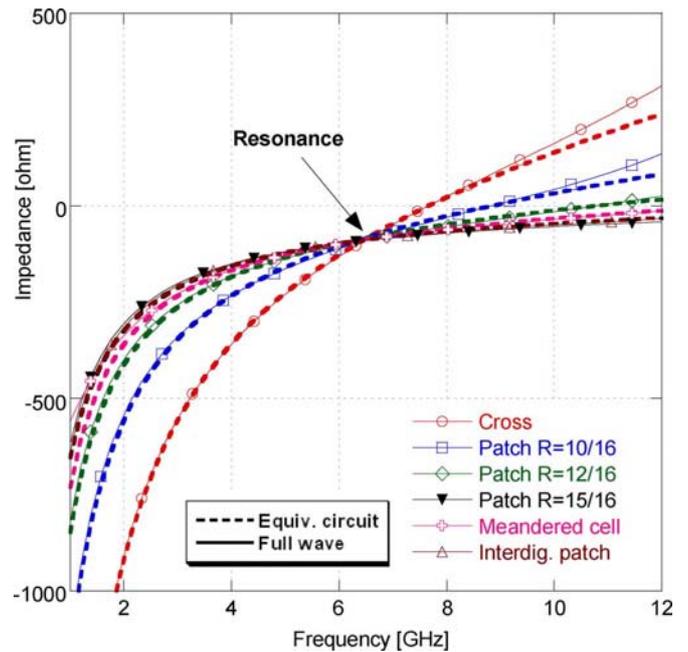

Fig. 6. Reactive impedance of the FSS of the different cell shapes.

TABLE I
INDUCTANCE AND CAPACITANCE VALUES OF THE FSSs UNDER ANALYSIS AND THE PERCENTAGE FRACTIONAL BANDWIDTH FBW$_\%$, COMPUTED BY THE RELATION (5)

|  | L [nH] | C [fF] | FBW[%] |
|---|---|---|---|
| Patch R=15/16, D=7.77mm | 0.311 | 242.96 | 19.35 |
| Interdig. patch, D=6.87 mm | 0.314 | 244.1 | 19.27 |
| Meander. cell, D=1.53 mm | 0.656 | 216.87 | 16.82 |
| Patch R=12/16, D=12.64 mm | 1.173 | 186.57 | 13.76 |
| Patch R=10/16, D=15.86 mm | 2.395 | 135.82 | 9.97 |
| Cross, D=13.25 mm | 5.353 | 80.77 | 6.06 |

Nevertheless, the simplified relation (1) correctly predicts the enlargement of the fractional bandwidth in the patch-type FSS when the distance between adjacent patches is increased and the periodicity is kept constant [13] (however, this case is not of practical interest since we do not obtain the resonance in correspondence with the same frequency). We also conclude that the bandwidth of the cross element is the smallest, whereas, among the patches, a higher ratio $R$ leads to a larger bandwidth of the FSS. The meandered structure is characterized both by a high capacitance, due to the strong coupling between adjacent lines, and a high inductance, due to the narrowness and the length of the lines. Indeed, these high values are instrumental for a substantial miniaturization of the cell in an AMC configuration. This small periodicity determines a total capacitance and a total inductance (depending on the periodicity also [7]) similar to other elements (see Table I).

Similar results for the bandwidth of crosses and patches were qualitatively explained in [14] by exploiting the theory of small antennas. The authors argued that the AMC bandwidth is reduced by the degree to which the fields are localized within the surface texture and do not uniformly fill its volume. Finally, it is worth mentioning that the patch-type FSS with very small gap is the more effective solution for blocking the TE surface waves. Indeed, the effectiveness of the FSS in blocking the normal magnetic field determines the TE mode cutoff frequency, and hence the TE bandwidth [15].

## IV. CONCLUSION

In this letter, a novel circuit approach for studying the bandwidth of HISs has been presented. As discussed in the body of this letter, the conventional representation of the HIS as a parallel *LC* circuit leads to an inaccurate estimate of the device bandwidth. For such a reason, the expression of the fractional bandwidth by using a more complete equivalent circuit, including also the series FSS inductance, has been derived. We have shown that the increase of the series inductance implies a reduction of the fractional bandwidth. It has therefore been demonstrated that the patch FSS element with a small gap between adjacent patches, even if strongly capacitive, is the best choice for designing a wideband AMC in correspondence with a fixed frequency. We have also shown that an extremely miniaturized HIS can be designed by employing a meandered shape element while maintaining wideband properties.